\documentclass[aps,twocolumn,groupedaddress,showpacs,prl,floatfix]{revtex4}
\usepackage{graphicx,rotating,subfigure,amsmath,amsfonts,amssymb,delarray}
\renewcommand{\vec}[1]{\boldsymbol #1}
\newcommand{\e}{\text{e}}
\newcommand{\im}{\text{i}}
\def\l{\left}
\def\r{\right}
\def\12{\frac{1}{2}}

\def\tr{\mbox{Tr}}

\begin{document}
% You should use BibTeX and apsrev.bst for references
\bibliographystyle{apsrev}

%\preprint{}

\title{Finite temperature fidelity susceptibility for one-dimensional quantum systems}
% Optional argument for running titles on pages
%\title[]{}


% other informatio
\author{J. Sirker}
% \email[]{jsirker@fkf.mpg.de}
%\thanks{}
\affiliation{Department of Physics and Research Center OPTIMAS, University of Kaiserslautern, D-67663 Kaiserslautern, Germany}
%%\affiliation{Max-Planck-Institute for Solid State Research, Heisenbergstr.~1,
%%  70569 Stuttgart, Germany}

\date{\today}

\begin{abstract}
  We calculate the fidelity susceptibility $\chi_f$ for the Luttinger
  model and show that there is a {\it universal} contribution linear
  in temperature $T$ (or inverse length $1/L$).  Furthermore, we
  develop an algorithm - based on a lattice path integral approach -
  to calculate the fidelity $F(T)$ in the {\it thermodynamic limit}
  for one-dimensional quantum systems. We check the Luttinger model
  predictions by calculating $\chi_f(T)$ analytically for free
  spinless fermions and numerically for the $XXZ$ chain.  Finally, we
  study $\chi_f$ at the two phase transitions in this model.
% Finally, we conjecture that
%   $\chi_f(T)\sim\ln^2 T_0/T$ at the isotropic Heisenberg point where
%   $T_0$ is a scale.
\end{abstract}
%%\pacs{05.10.Cc}
%%\pacs{05.70.-a}
%\keywords{}

\maketitle

%% \section{Introduction}
%% \label{Intro}
Phase transitions are usually identified by considering suitably
defined order parameters. Lately, new concepts originating from
quantum information theory have been put forward which allow to detect
phase transitions without any prior knowledge of the order parameter
\cite{VidalLatorre,VenutiZanardi,SchwandtAlet,ZhouOrus,ZanardiPaunkovic,ZanardiVenuti,ZanardiQuan,ChenWang,YouLi,Yang,Dillenschneider,Sarandy}.
The most widely used measures are the {\it entanglement entropy}
\cite{VidalLatorre} and the {\it fidelity}
\cite{VenutiZanardi,SchwandtAlet,ZhouOrus,ZanardiPaunkovic,ZanardiVenuti,ZanardiQuan,ChenWang,YouLi,Yang}.
The latter approach is based on the notion that at a quantum phase
transition the ground state wave function is expected to change
dramatically with respect to a parameter $\lambda$ driving the
transition \cite{ZanardiPaunkovic}. If the Hamiltonian is given by
$\hat{H}_\lambda=\hat{H}_0 +\lambda \hat{O}$, then the fidelity is
defined as
\begin{equation}
\label{Fidelity1}
F_0(\lambda)=\sqrt{\langle\Psi_0|\Psi_\lambda\rangle\langle\Psi_\lambda|\Psi_0\rangle/\langle\Psi_0|\Psi_0\rangle\langle\Psi_\lambda|\Psi_\lambda\rangle}
\end{equation}
where $|\Psi_0\rangle$ [$|\Psi_\lambda\rangle$] is the ground state
wave function of $\hat{H}_0$ [$\hat{H}_\lambda$], respectively. The
fidelity has been studied analytically for one-dimensional (1D) models
like the transverse Ising or the $XY$ model
\cite{ZanardiPaunkovic,ChenWang,ZanardiQuan} as well as numerically
for a number of other systems
\cite{VenutiZanardi,ChenWang,SchwandtAlet}.  Importantly, the fidelity
approach connects many different areas of physics and is not
restricted to the study of phase transitions. The overlap between wave
functions also plays a central role for scattering problems
(Anderson's orthogonality catastrophy) \cite{AndersonOC}, as a measure
for variational wave functions, for quantum information processing
\cite{RyanEmerson}, the Loschmidt echo \cite{LesovikHassler}, and for
quench dynamics \cite{deGrandiGritsev}.
% If parameters of the Hamiltonian are suddenly quenched - a technique
% often used experimentally in cold atomic gases - it has been shown
% that the probability to excite the system away from its ground state
% is proportional to the fidelity.
Apart from calculating the fidelity for specific models it is
therefore of great interest to understand possible {\it universal
  behavior}. For critical 1D quantum systems such universality is
often related to conformal invariance. Important examples are the
scaling of the free energy \cite{Affleck86} and the entanglement
entropy \cite{VidalLatorre} with system size $L$ and temperature $T$.

In this letter we will introduce a new finite temperature (mixed
state) fidelity and show that it leads to the fidelity susceptibility
$\chi_f$ used in recent quantum Monte Carlo simulations
\cite{SchwandtAlet}. We then show that $\chi_f$ for the Luttinger
model has a {\it universal} term linear $T$.  Similarly, there is a
universal term $\sim 1/L$ for a finite system at zero temperature.
$\chi_f(T=0)$ in the thermodynamic limit, on the other hand, depends
on a cutoff, a fact, which has been missed in an earlier work
\cite{Yang}.  Furthermore, we express $F(T)$ in the {\it thermodynamic
  limit} for any 1D quantum system as a function of the largest
eigenvalues of three transfer matrices. This allows for a very
efficient numerical calculation of the fidelity making it an ideal
tool for finding phase transitions without any prior knowledge of the
order parameters. We apply this method to study $\chi_f(T)$ for the
$S=1/2$ $XXZ$ chain with respect to a small change in the anisotropy
$\Delta$ allowing us to check our results for the Luttinger model
directly. A further check is provided by an analytic calculation of
$\chi_f(T)$ in the free fermion case. Finally, we extract
$\chi_f(T=0)$ for the $XXZ$ model from the numerical data and discuss
its behavior at the two critical points.

We can generalize (\ref{Fidelity1}) to finite temperatures so that
$F_T(0)=1$ and $\lim_{T\to 0}F_T(\lambda) = F_0(\lambda)$ by
\begin{equation}
\label{Fidelity2}
F_T(\lambda)=\sqrt{\tr \{\e^{-\beta\,\hat{H}_0/2}\e^{-\beta\,\hat{H}_\lambda/2}\}}/(Z_0\, Z_\lambda)^{1/4}
%% /\sqrt{\tr \e^{-\beta \hat{H}_0} \tr\e^{-\beta \hat{H}_\lambda}}}
\end{equation}
where $\beta =1/T$, $Z_0=\tr\,\e^{-\beta \hat{H}_0}$,
and $Z_\lambda=\tr\,\e^{-\beta \hat{H}_\lambda}$.
% Note that this generalization used here is
% different from the mixed state fidelity defined in \cite{ZanardiQuan}.
For a many-body system the fidelity is expected to vanish
exponentially with the number of particles $N$ no matter how small the
driving parameter $\lambda$ is \cite{AndersonOC}. 
%% This behavior is related to Anderson's
%% orthogonality catastrophe \cite{AndersonOC}.  
The fidelity density $f(\lambda)=-\frac{1}{N}\ln F$, however, stays
finite.  
%% Furthermore, we want to consider a small change of the tuning
%% parameter $\lambda$.  
Since $f(\lambda = 0)=0$ is a minimum, the first term in an expansion
for small $\lambda$ vanishes giving rise to the definition of the
fidelity susceptibility $\chi_f = (\partial^2 f/\partial
\lambda^2)_{\lambda=0}$ \cite{YouLi}. 
% $\chi_f$ is an important
% quantity because it is directly proportional to the propability to
% excite the system away from the ground state following a quantum
% quench with a small quench amplitude $\lambda$ \cite{deGrandiGritsev}.
From Eq.~(\ref{Fidelity2}) we find that
\begin{equation}
\label{Fidelity3}
\chi_f=\frac{1}{N}\int_0^{\beta/2} \tau\,d\tau \left\{\langle \mathcal{T}\hat{O}(\tau)\hat{O}(0)\rangle -\langle\hat{O}\rangle^2\right\} 
\end{equation}
where $\mathcal{T}$ denotes time ordering and
$\hat{O}(\tau)=\exp(\tau\hat{H}_0)\hat{O}\exp(-\tau\hat{H}_0)$. In the
following, we will consider the case $\hat{O}({\tau}) =\sum_r
\hat{o}(r,\tau)$ where $\hat{o}(r,\tau)$ is a local operator. By using
a Lehmann representation, Eq.~(\ref{Fidelity3}) can be shown to be
consistent for $T\to 0$ with the ground state
fidelity directly obtained from the definition (\ref{Fidelity1})
\cite{YouLi}. 
% This can be shown by a Lehmann representation which
% yields $\chi_f(T\to 0)=N^{-1}\sum_{m\neq 0}|\langle
% \Psi_0|\hat{O}|m\rangle |^2/(E_0-E_m)^2$ where $|m\rangle$ is an
% excited state and $E_m$ the corresponding energy.
Eq.~(\ref{Fidelity3}) has previously been used to {\it define}
$\chi_f(T)$ \cite{SchwandtAlet}. Here this expression
for $\chi_f(T)$ in terms of a correlation function directly follows
from Eq.~(\ref{Fidelity2}).  Note, however, that $F(T)$ in
(\ref{Fidelity2}) is different from the mixed state fidelity as
defined in \cite{ZanardiQuan,ZanardiVenuti} which does not allow to
express the corresponding $\chi_f$ as a simple correlation function.
Importantly, it has been shown that if $\chi_f$ as obtained from the
mixed state fidelity in \cite{ZanardiQuan,ZanardiVenuti} diverges then
so does $\chi_f$ as given in (\ref{Fidelity3}) and vice versa
\cite{SchwandtAlet}. Finally, we note that if
$[\hat{H}_0,\hat{O}]=0$ then $\chi_f(T) = \chi/8T$ with
$\chi=\langle(\sum_r\hat{o}_r)^2\rangle/(NT)$ being the regular
susceptility.

The generic low-energy effective theory for a gapless 1D quantum
system is the Luttinger model \cite{GiamarchiBook}
\begin{equation}
\label{LL1}
H_{LL}=\frac{v}{2}\int_{-L/2}^{L/2} dx \l[\frac{K}{2}\Pi^2 +\frac{2}{K}(\partial_x\phi)^2\r] \; .
\end{equation}
Here $v$ is a velocity, $L=Na$ the length with $a$ being the lattice
constant, and $K$ the Luttinger parameter. $\phi$ is a bosonic field
obeying the standard commutation rule
$[\phi(x),\Pi(x')]=\im\delta(x-x')$ with $\Pi=\im
v^{-1}\partial_\tau\phi$. In general, {\it both} $K$ and $v$ will
change as a function of a driving parameter $\lambda$ in the
Hamiltonian of the microscopic model.

The operator appearing in (\ref{Fidelity3}) is therefore given by
$\hat{O}=\hat{O}_1+\hat{O}_2$ with
\begin{equation}
\label{LL2}
\hat{O}_{1,2} = \frac{\alpha_{1,2}}{2}\int_{-L/2}^{L/2}\, dx\,\l(\frac{K}{2}\Pi^2\pm\frac{2}{K}(\partial_x\phi)^2\r) 
\end{equation}
% \begin{eqnarray}
% \label{LL2}
% \hat{O}_1 &=& \frac{\partial v/\partial\lambda}{2}\l(\frac{K}{2}\Pi^2+\frac{2}{K}(\partial_x\phi)^2\r) \nonumber \\
% \hat{O}_2 &=& \frac{v}{2K}\frac{\partial K}{\partial\lambda}\l(\frac{K}{2}\Pi^2-\frac{2}{K}(\partial_x\phi)^2\r)
% \end{eqnarray}
and $\alpha_{1}= \partial v/\partial\lambda$, $\alpha_2=v(\partial
K/\partial\lambda)/K$. 
% We want to emphasize that while we specialize
% later to the $XXZ$ model, the results here are generic for any model
% described at low temperatures by (\ref{LL1}) and $\lambda$ can
% represent any microscopic parameter changing $v$, $K$ in the effective
% model.
We note that $\hat{O}_1$ is proportional to the Hamiltonian itself.
By rescaling $\Pi\to\sqrt{2/K}\Pi$, $\phi\to\sqrt{K/2}\phi$ we can
express the Hamiltonian and therefore also $\hat{O}_1$ as the sum of
the holo- and antiholomorphic components of the energy-momentum tensor
\cite{Lukyanov}. %% $T$ ($\bar{T}$), respectively \cite{Lukyanov}.
The finite temperature correlation function (\ref{Fidelity3})
involving $\hat{O}_1$ can then be calculated with the help of the
operator product expansion for this conformally invariant theory.
While the cross term vanishes, the integral (\ref{Fidelity3}) for the
operator $\hat{O}_2$ is divergent and we introduce a cutoff by
replacing $\int_0^{\beta/2}\to\int_{\tau_0}^{\beta/2}$. Combining both
contributions we find in the thermodynamic limit at low temperatures
\begin{equation}
\label{LL3}
\chi_f(T)=\frac{\Lambda}{8K^2}\l(\frac{\partial K}{\partial\lambda}\r)^2 + \frac{\pi c}{24v^3}\l(\frac{\partial v}{\partial\lambda}\r)^2  T \; .
\end{equation}
with $\Lambda = 1/(\pi v\tau_0)$ and $c=1$ being the {\it central
  charge} of the free bosonic model. The universality found here for
the leading linear temperature dependence of $\chi_f$ is reminiscent
of the universal term in the free energy of 1D critical
quantum systems quadratic in temperature \cite{Affleck86}. We also
want to remark that a universal subleading term in the zero
temperature fidelity has recently been discovered in certain systems
\cite{VenutiSaleur}.  
% Note, however, that the universality here is
% based on the fact that the operator $\hat{O}$ has a part proportional
% to the energy-momentum tensor and is therefore not as generic as the
% occurrence of a universal term $\sim T^2$ in a $1+1$ dimensional
% conformal field theory \cite{Affleck86}. 

$\chi_f(T=0)$ as obtained in (\ref{LL3}), on the other hand, is
cutoff dependent. This seems to be in contrast to an earlier work
\cite{Yang} where $\chi_f$ was directly calculated at zero temperature
using the definition (\ref{Fidelity1}).  This leads to
$\chi_f=(\partial K/\partial \lambda)^2/(4NK^2) \sum_{k>0}$ and the
result in \cite{Yang} is obtained if one assumes $N/2$ $k$-values in
the sum. The Luttinger model, however, is a continuum model and the
sum therefore not restricted. If we introduce a UV cutoff $N\Lambda/2$
then the first term in (\ref{LL3}) is reproduced.

Similarly, we can calculate $\chi_f$ for the Luttinger model of finite
size $L$ at zero temperature using Eq.~(\ref{Fidelity3}). Due to the
unusual imaginary-time integration the result cannot be obtained by
simply replacing $v/T\to L$ but rather the second term in (\ref{LL3})
gets replaced by $c(\partial v/\partial \lambda)^2/(8v^2L)$.

\begin{figure}
\includegraphics*[width=0.99\columnwidth]{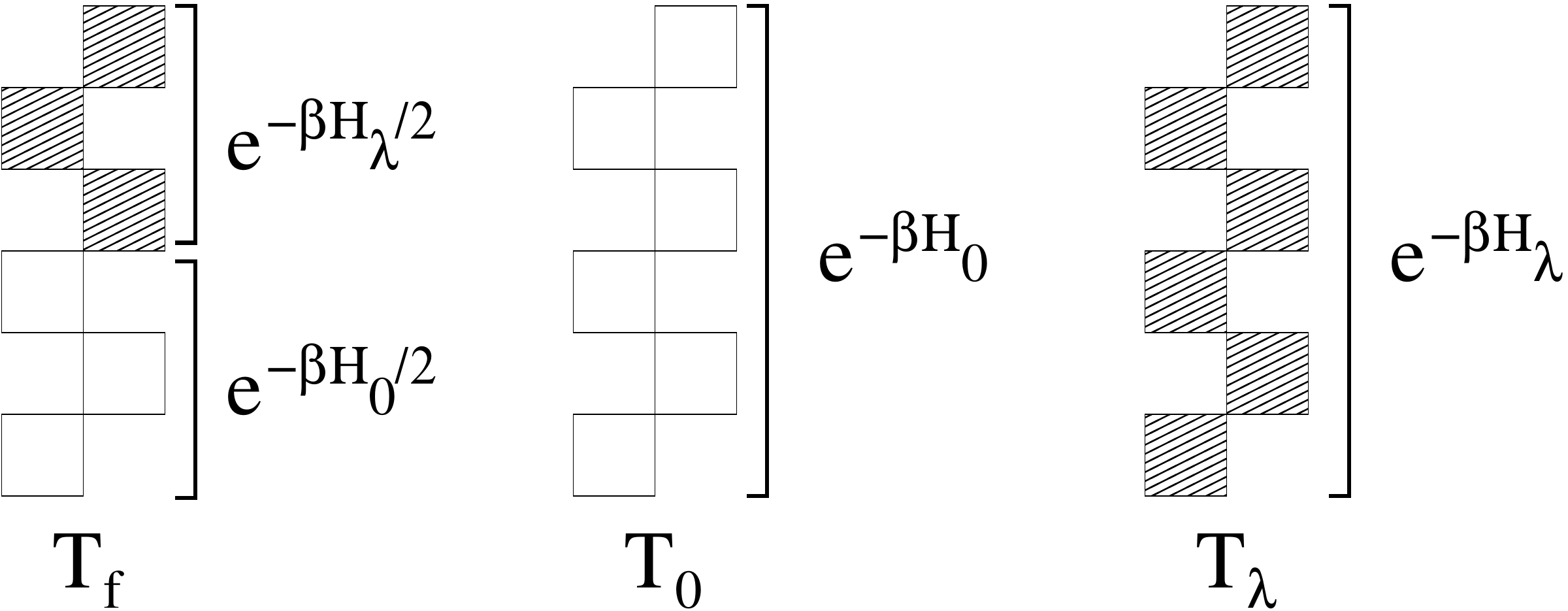}
\caption{Transfer matrices for calculating $F(T)$. Each open [shaded]
  plaquette represents a local Boltzmann weight $\exp(-\epsilon
  h_0^{r,r+1})$ [$\exp(-\epsilon h_\lambda^{r,r+1})$], respectively,
  with $\epsilon$ being the Trotter parameter.}
\label{TM.fig}
\end{figure}

By using a lattice path integral representation, a 1D quantum model
can be mapped onto a two-dimensional classical model with the
additional dimension corresponding to the inverse temperature. For the
fidelity (\ref{Fidelity2}) this amounts to separate Trotter-Suzuki
decompositions for each of the exponentials.  We consider a
Hamiltonian with nearest-neighbor interaction
% (or short
% range interactions which can be rewritten as nearest-neighbor
% interactions by enlarging the unit cell)
and decompose the Hamiltonian into
$H_{0,\lambda}^e=\sum_{r\,\rm{even}}h_{0,\lambda}^{r,r+1}$ and
$H_{0,\lambda}^o=\sum_{r\,\rm{odd}}h_{0,\lambda}^{r,r+1}$.  This
allows us to write $\exp(-\beta H_0) =
\lim_{M\to\infty}[\exp(-\epsilon H_0^e)\exp(-\epsilon H_0^o)]^M$ and
equivalently for the other exponentials in (\ref{Fidelity2}). Here
$\epsilon=\beta/M$ is the Trotter parameter. Rearranging the local
Boltzmann weights we can define the column transfer matrices
depicted in Fig.~\ref{TM.fig}.
The spectra of these transfer matrices have a gap between the largest
and the next-leading eigenvalue thus allowing it to perform the
thermodynamic limit exactly \cite{SirkerKluemperEPL}. For the fidelity
density we find
\begin{equation}
\label{Fidelity4}
f_T(\lambda)=-\frac{1}{N}\ln F = -\frac{1}{4}\ln\left(\frac{\Lambda_f}{\sqrt{\Lambda_0\Lambda_\lambda}}\right)
\end{equation}
where $\Lambda_f$, $\Lambda_0$, and $\Lambda_\lambda$ are the largest
eigenvalues of the transfer matrices $T_f$, $T_0$, and $T_\lambda$
defined in Fig.~\ref{TM.fig}, respectively. Because $f_T(0)=\partial
f_T/\partial\lambda|_{\lambda=0}=0$ we can calculate the fidelity
susceptibility by $\chi_f(T)=2\lim_{\lambda\to
  0}f_T(\lambda)/\lambda^2$, i.e., without having to resort to
numerical derivatives. The transfer matrices can be efficiently
extended in imaginary time direction - corresponding to a successive
reduction in temperature - by using a density-matrix renormalization
group algorithm applied to transfer matrices (TMRG). If we are mainly
interested in $\chi_f$ then only small parameters $\lambda$ have to be
considered, allowing it to renormalize all three transfer matrices
with the same reduced density matrix. Apart from the two different
Boltzmann weights necessary to form the three transfer matrices
depicted in Fig.~\ref{TM.fig} the algorithm can therefore proceed in
exactly the same way as the TMRG algorithm to calculate thermodynamic
quantities.  For technical details of the algorithm the reader is
therefore referred to
Refs.~\cite{BursillXiang,SirkerKluemperEPL}.

\begin{figure}
\includegraphics*[width=0.9\columnwidth]{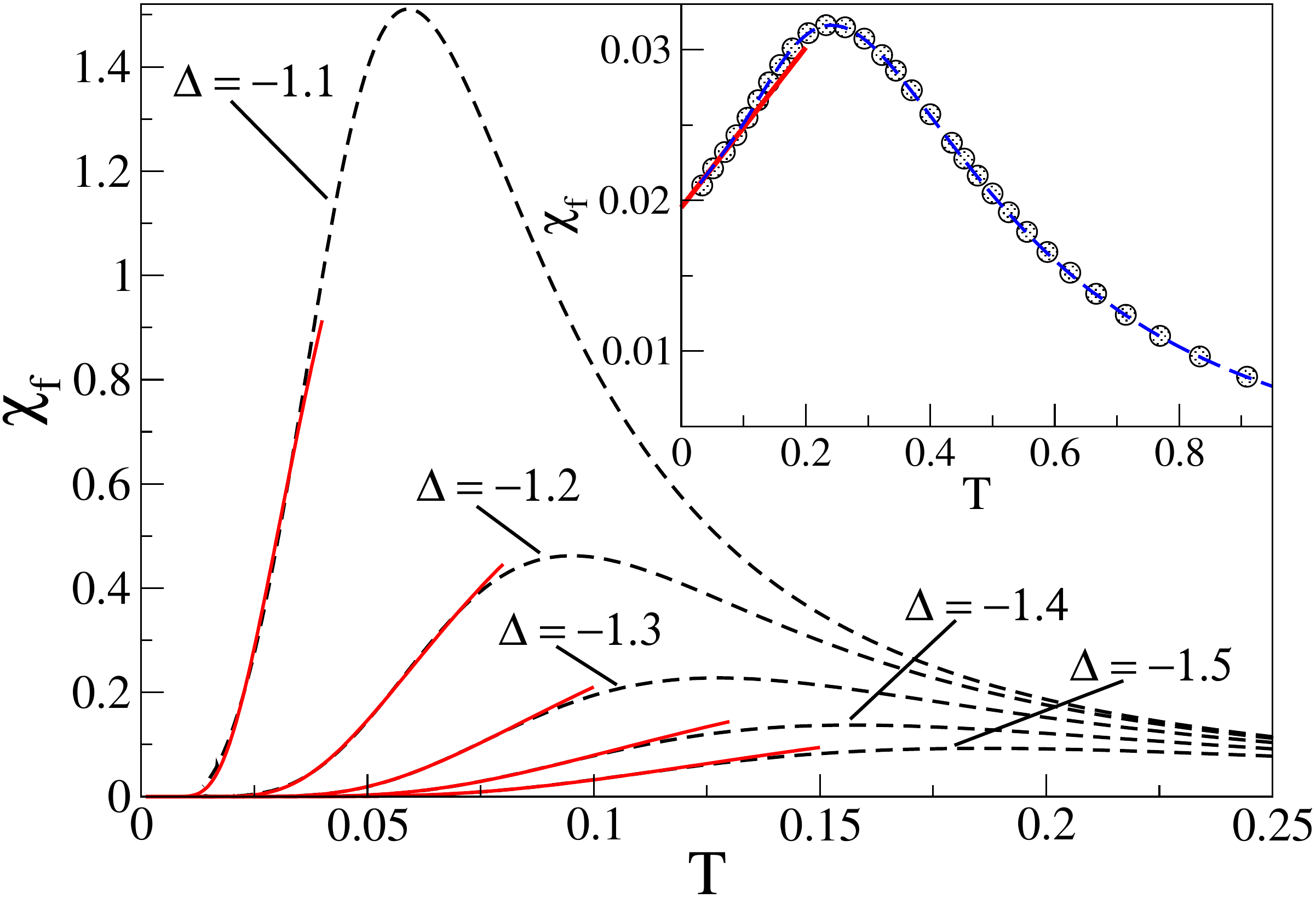}
\caption{TMRG data (dashed) and fits (solid lines) for $\chi_f(T)$ in
  the ferrmomagnetic Ising regime. Inset: TMRG data for the free
  fermion case (circles) compared to the exact solution (dashed line)
  and the low-$T$ asymtotics (solid line).}
\label{TM.fig2}
\end{figure}

In the following, we want to study $\chi_f(T)$ for the $XXZ$ model defined by
\begin{equation}
\label{XXZ}
H=J\sum_r\left\{S^x_rS^x_{r+1}+S^y_rS^y_{r+1}+\Delta S^z_rS^z_{r+1}\right\} 
\end{equation}
with respect to a change in anisotropy $\Delta$. Here $\vec{S}$ is a
spin $S=1/2$ operator and $J$ the exchange constant which we set to
$1$. We note that $\chi_f$ at zero temperature for finite chains has
previously been studied in \cite{VenutiZanardi,ChenWang}. The model is
gapless for $-1\leq\Delta\leq 1$ and gapped for $|\Delta|>1$. At
$\Delta = 0$ the model describes non-interacting spinless fermions and
$\chi_f$ can be calculated exactly. The various diagrams can be
combined into two contributions
 \begin{eqnarray}
 \label{fF1}
 \chi_f^{(1)}&\!\!\!\!\! =& \!\!\!\!\!\frac{1}{4\pi^3}\!\int_{-\pi}^{\pi}\!\!\!\!\! dk_1\, dk_2\, dk_3 \frac{1-\e^{-\beta x/2}}{x^2}y\bar{n}^F_{k_1}\bar{n}^F_{k_2}n^F_{k_3}n^F_{k_1+k_2-k_3} \nonumber \\
\chi_f^{(2)} &\!\!\! = & \!\!\!\frac{1}{16\pi^3T^2}\left[\int_{-\pi}^\pi\!\!\!\!\! dk\,\cos k\, n^F_k\right]^2\int_{-\pi}^\pi\!\!\!\!\! dk\,\cos^2 k\, n^F_k \,\bar{n}^F_k
%% \chi_f^{(2)} &=& \frac{1}{4\pi^3T^2}\int_{-\pi}^{\pi} dk\, \cos^2 k\, n^F_k\, \bar{n}^F_k 
 \end{eqnarray}
% \begin{widetext}
% \begin{eqnarray}
% \label{fF1}
% \chi_f^{(1)}&=& \frac{1}{4\pi^3}\int_{-\pi}^{\pi}\! dk_1\, dk_2\, dk_3 \frac{1-\e^{-\beta x/2}}{x^2}[\cos^2(k_1-k_3)-\cos(k_1-k_3)\cos(k_2-k_3)](1-n^F_{k_1})(1-n^F_{k_2})n^F_{k_3}n^F_{k_1+k_2-k_3} \nonumber \\
% \chi_f^{(2)} &=& \frac{1}{4\pi^3T^2}\int_{-\pi}^{\pi} dk\, \cos^2 k\, n^F_k(1-n^F_k) 
% \end{eqnarray}
% \end{widetext}
 with $x=\cos k_1 +\cos k_2 -\cos k_3 -\cos(k_1+k_2-k_3)$,
 $y=\cos^2(k_1-k_3)-\cos(k_1-k_3)\cos(k_2-k_3)$, $n^F_k =
 1/[1+\exp(\beta\cos k)]$ and $\bar{n}^F_k=1-n^F_k$. The first
 contribution at low temperatures yields $\chi_f^{(1)}= 0.19537(\pm
 5) +\mathcal{O}(T^2)$ whereas the second is given by $\chi_f^{(2)}
 =T/(6\pi)$. In the inset of Fig.~\ref{TM.fig2} the exact solution for
 $\Delta = 0$ is compared with the TMRG data obtained from $\chi_f =
 2f(\Delta+\delta\Delta)/(\delta\Delta)^2$ with $\delta\Delta =
 10^{-3}$.
The relative error without any extrapolation is less than $0.1\%$ for
$T>0.1$ and less than $1\%$ for $T>0.04$.

In the gapped regime, $|\Delta|>1$, the fidelity susceptibility will
show activated behavior. Following the arguments in
\cite{JohnsonMcCoy} for the magnetic susceptibility we expect
$\chi_f\sim T^{-3/2}\exp(-\gamma/T)$ with $\gamma = -\Delta -1$ being
the spectral gap for $\Delta < -1$ and $\gamma$ being half the
spectral gap for $\Delta>1$. Note that in the latter case the spectral
gap is exponentially small for $\Delta\gtrsim 1$ making it difficult
to detect numerically. As shown in Fig.~\ref{TM.fig2} a fit of the
data for $\Delta <-1$ is consistent with this scaling form with fitted
$\gamma$ values close to the one theoretically expected.
% The fitted $\gamma$ values for $\Delta =-1.1,\, 1.2$ are about $10\%$
% larger than expected theoretically which we attribute to the presence
% of corrections $\sim\exp(-2\gamma/T)$ to the scaling formula (??).

\begin{figure}
\includegraphics*[width=0.9\columnwidth]{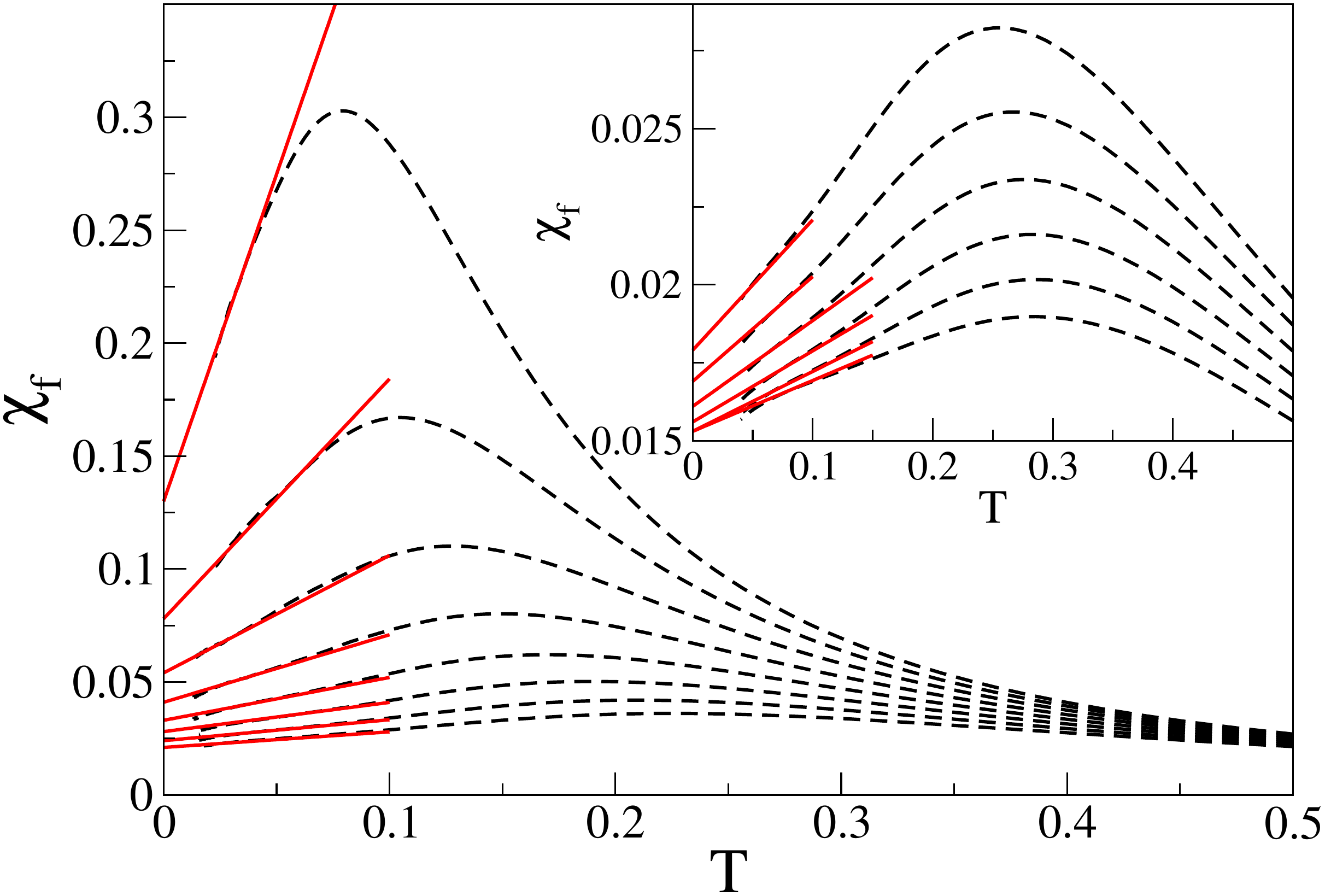}
\caption{TMRG data (dashed) and low temperature fits (solid lines) for
  $\chi_f$ with the slope fixed by (\ref{LL3}). The main figure
  (inset) shows data for $\Delta=-0.8,-0.7,\cdots,-0.1$
  ($\Delta=0.1,0.2,\cdots,0.6$) from top to bottom, respectively.}
\label{TM.fig3}
\end{figure}

\begin{figure}
\includegraphics*[width=0.9\columnwidth]{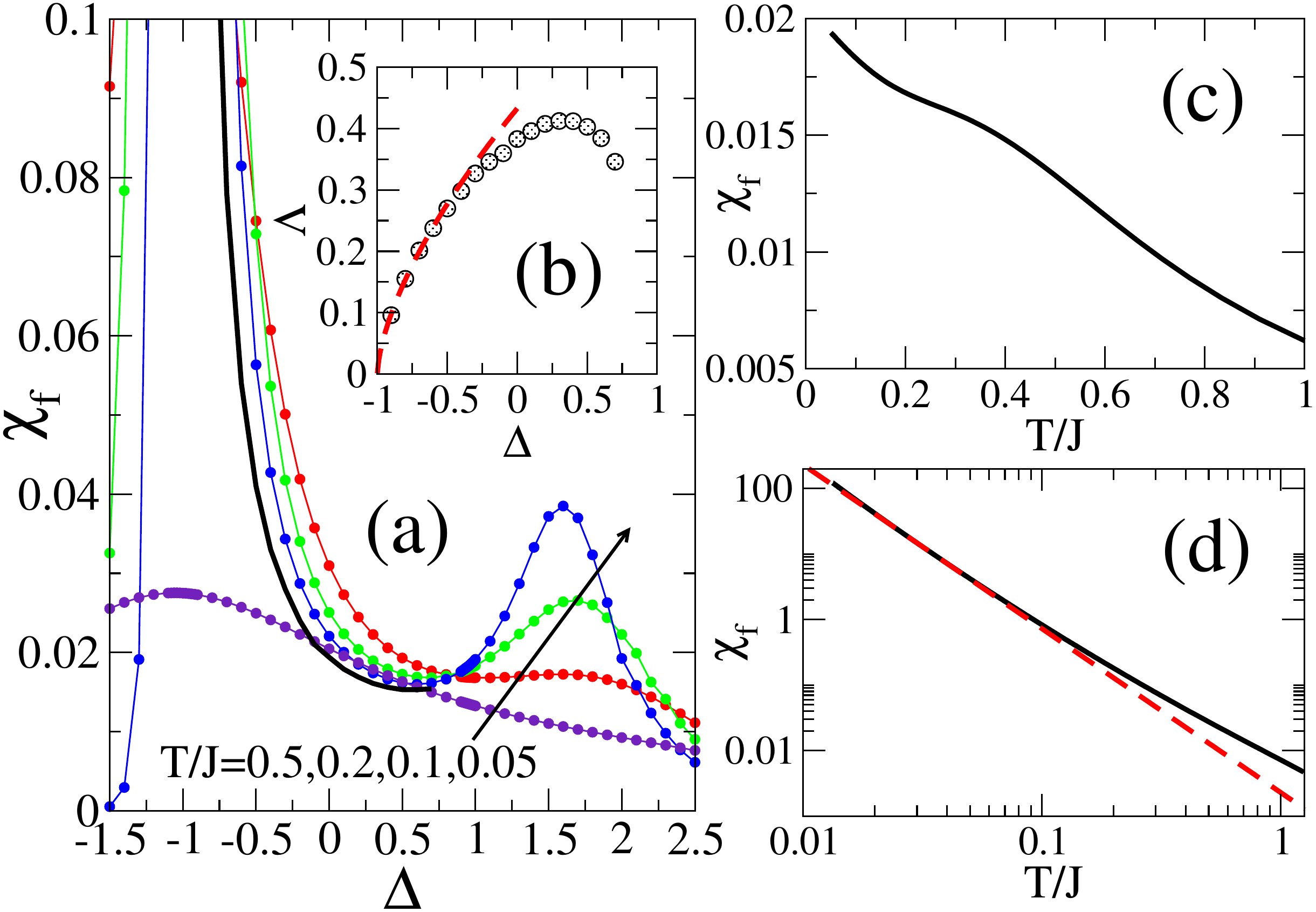}
\caption{(a) The solid line depicts the extrapolated $T=0$ curve in
  the critical regime, the symbols the finite temperature curves as
  indicated on the plot. (b) The momentum cutoff $\Lambda$ as a
  function of $\Delta$ (symbols).  The dashed line shows a fit for
  $\Delta\gtrsim -1$. (c) $\chi_f(T)$ for $\Delta=1$, and (d) for
  $\Delta=-1$ with the dashed line representing a low-$T$ fit
  as discussed in the text.}
\label{TM.fig4}
\end{figure}

In the gapless regime, $-1<\Delta\leq 1$, we know from the Bethe
ansatz that $K=\pi/(\pi-\arccos\Delta)$ and
$v=\pi\sqrt{1-\Delta^2}/(2\arccos\Delta)$. This allows us to check the
universality of the contribution linear in $T$ in (\ref{LL3}) by a
direct comparison with the TMRG data (see Fig.~\ref{TM.fig3}) and to
accurately extract $\chi_f(T=0)$.
This method fails, however, close to $\Delta =1$ where corrections due
to Umklapp scattering start to become important (as will be discussed
in more detail below) as well as very close to $\Delta=-1$ where the
Luttinger model fails because the dispersion of the elementary
excitations becomes quadratic. The fidelity susceptibility as a
function of $\Delta$ for various temperatures as well as the
extrapolated $T=0$ curve are shown in Fig.~\ref{TM.fig4}(a).
Comparing with the theoretical result (\ref{LL3}) we can also extract
the momentum cutoff $\Lambda$ (see Fig.~\ref{TM.fig4}(b)). There is a
clear divergence of $\chi_f$ at the first order phase transition
$\Delta = -1$.  A fit of the extrapolated zero temperature curve shown
in Fig.~\ref{TM.fig4}(a) gives $\chi_f(\Delta\gtrsim -1)\sim
0.017/(\Delta + 1)^{1.26}$. This requires that the cutoff $\Lambda$
vanishes for $\Delta\to -1$ because otherwise we would find from
(\ref{LL3}) a divergence $\sim 1/(\Delta+1)^{2}$ as predicted in
\cite{Yang}. Indeed, a fit of the extracted momentum cutoff as shown
in Fig.~\ref{TM.fig4}(b) yields $\Lambda\sim 0.43(\Delta+1)^{0.65}$
and therefore $\chi_f(\Delta\gtrsim -1)\sim 0.013/(\Delta + 1)^{1.35}$
which is consistent with the direct fit.

At the Kosterlitz-Thouless (KT) phase transition, $\Delta=1$, on the
other hand, the behavior is different. Here the finite temperature
data show that a maximum in $\chi_f$ at $\Delta>1$ exists which shifts
to smaller $\Delta$ with decreasing temperature. The dependence of the
cutoff $\Lambda$ near $\Delta =1$ seems to be consistent with
$\Lambda\sim\Lambda_1 + (1-\Delta)^\alpha$ with a constant $\Lambda_1$ and
an exponent $\alpha$ {\it both} greater than zero. If this is indeed
the case, we find from (\ref{LL3}) that $\chi_f(\Delta\lesssim 1)\sim
\Lambda_1/[16\pi^2(1-\Delta)]$.

Finally, we want to discuss the temperature dependence of $\chi_f$
right at the phase transitions. For $\Delta=-1$, shown in
Fig.~\ref{TM.fig4}(d), we find a divergence $\chi_f\approx 0.002(\pm
1)T^{-2.5(\pm 1)}$ where the error is determined by a variation of the
fit interval. As argued above, we also expect $\chi_f(T)$ to diverge
for $T\to 0$ and $\Delta =1$. The numerical data, shown in
Fig.~\ref{TM.fig4}(c), however, do not easily allow to extract the low
temperature behavior. If we assume that $\lim_{\Delta\to 1}\Lambda =
\Lambda_1 >0$ then we can calculate the temperature dependence
analytically as follows. At the isotropic point, Umklapp scattering is
marginally irrelevant and the Luttinger
parameter %% for $\Delta\lesssim 1$
has to be replaced by a running coupling constant $K\to
1+g_\parallel(l)/2$ where $l=\ln T_0/T$ with a scale $T_0$ of order
$J$ and $g_\parallel(l) = (2K^*-2)/\tanh[(2K^*-2)l+{\rm
  atanh}((2K^*-2)/g_\parallel^0)]$ \cite{GiamarchiBook}. For $T=T_0$
we have $l=0$ and $g_\parallel(0)=g_\parallel^0$ while for $T\to 0$ it
follows that $l\to\infty$ and $K=1+g_\parallel/2\to K^*$ where $K^*$
is the fix point value. For $l$ large we can neglect the part $\propto
g_\parallel^0$. We therefore obtain
\begin{eqnarray}
\label{LL4}
\chi_f(\Delta=1,T)&=&\lim_{\Delta\to 1}\frac{\Lambda_1}{32(1+g_\parallel/2)^2}\l(\frac{\partial g_\parallel}{\partial\Delta}\r)^2\nonumber \\
&\stackrel{T\ll T_0}{\to}& \frac{2\Lambda_1}{9\pi^4}\ln^2(T_0/T)\; .
\end{eqnarray}
While this prediction %% based on a seemingly obvious assumption
resolves some confusion about the behavior of $\chi_f$ at a KT
transition \cite{VenutiZanardi,YouLi,ChenWang,Yang} it cannot be
reliably tested by comparing with the TMRG data.  While the term in
(\ref{LL4}) should dominate at very low temperatures, subleading
corrections might be of equal importance in the temperature range
accessible numerically.

To summarize, we have shown that the fidelity susceptibility for the
Luttinger model has a universal term linear in temperature or inverse
length. Apart from being relevant for quantum phase transitions we
believe that this result is also important to analyze sudden quantum
quenches.
% This is related to the fact that a part of the considered
% operator in this case is proportional to the energy-momentum tensor.
Furthermore, we have introduced a numerical method 
%- based on a transfer matrix approach - 
to calculate the finite temperature fidelity in the thermodynamic
limit for any 1D quantum system with short range interactions.
Finally, based on a RG treatment, we have predicted a $\ln^2T$
divergence of $\chi_f$ at the KT transition in the $XXZ$ model.

%% We have applied this method to the $XXZ$
%% model and tested the Luttinger model predictions. 
% A partly open
% problem remains the behavior of $\chi_f(T)$ for the isotropic
% antiferromagnet.  Theoretically, we expect a divergence
% $\sim\ln^2(T_0/T)$ but have not been able to numerically confirm this
% behavior due to the inability to reach low enough temperatures.

\begin{acknowledgments}
  JS thanks I. Affleck and F. Alet for valuable discussions. This work
  was supported by the MATCOR school of excellence.
\end{acknowledgments}

\end{document}